# Generating Synthesized Computed Tomography (CT) from Cone-Beam Computed Tomography (CBCT) using CycleGAN for Adaptive Radiation Therapy

Xiao Liang[§], Liyuan Chen[§], Dan Nguyen, Zhiguo Zhou, Xuejun Gu, Ming Yang, Jing Wang, Steve Jiang

Medical Artificial Intelligence and Automation Laboratory, Department of Radiation Oncology, University of Texas Southwestern Medical Center, Dallas, TX, USA

[§]Co-first authors.

E-mails: Steve.Jiang@UTSouthwestern.edu and Jing.Wang@UTSouthwestern.edu

**Abstract**

Cone beam computed tomography (CBCT) images can be used for dose calculation in adaptive radiation therapy (ART). The main challenges are the large artefacts and inaccurate Hounsfield unit (HU) values. Currently, deformed planning CT images are often used for this purpose, although anatomical accuracy might be a concern. Ideally, we would like to convert CBCT images to CT images with artifacts removed or greatly reduced and HU values corrected while keeping the anatomical accuracy. Recently, deep learning has achieved great success in image-to-image translation tasks. It is very difficult to acquire paired CT and CBCT images with exactly matching anatomy for supervised training. To overcome this limitation, we developed and tested a cycle generative adversarial network (CycleGAN) which is an unsupervised learning method and does not require paired training datasets to synthesize CT images from CBCT images. The synthesized CT (sCT) images have been compared with the deformed planning CT (dpCT) showing visual and quantitative similarity with artifacts being removed and HU value errors being reduced from 71.78 HU to 27.98 HU. Dose calculation accuracy using sCT images has been improved over the original CBCT images, with the average Gamma Index passing rate increased from 95.4% to 97.4% for 1 mm/1% criteria. A deformable phantom study has been conducted and demonstrated better anatomical accuracy for sCT over dpCT.

## 1. Introduction

Inter-factional anatomical change occurs to many cancer patients under radiotherapy and to many tumor sites such as head and neck (H&N) cancer, often due to tumor shrink or weight loss during treatment course. Only relying on computed tomography (CT) images acquired before treatment may increase the risk of tumor underdose and organs at risk (OARs) overdose. Several studies have shown compromised tumor coverage caused by anatomical changes during treatment courses. One pilot study on H&N anatomical changes showed that gross tumor volume (GTV) decreased at a median rate of 1.8% per treatment day and a median total relative loss of 69.5% on the last day of treatment with a median mass displacement of 3.3 mm. Also, the parotid gland showed a median reduction in volume of 0.19 cm$^3$ per day and a median shift of 3.1 mm (Barker *et al.*, 2004). Another study showed that H&N patients with significant anatomical changes like neck diameter reduction greater than 10% or weight loss more than 5% would be more likely to suffer from deleterious side effects such as xerostomia (You *et al.*, 2012). A retrospective study that included 13 H&N patients showed that the doses to 95% (D95) of the planning target volume (PTV) were reduced in 92% of patients. Also, the maximum dose (Dmax) of the spinal cord increased in all patients and the Dmax to brainstem increased in 85% of patients because of anatomical changes (Hansen *et al.*, 2006).

Findings from clinical trials have indicated that adaptive radiation therapy (ART) improves the dosimetric quality of radiation therapy plans in H&N patients. A clinical trial evaluating ART benefits on H&N patients was conducted by acquiring daily in-room CT-on-rails or cone beam computed tomography (CBCT) images to track anatomical changes, and then using deformable image registration (DIR) to align contours on new images for replanning, if needed (Schwartz *et al.*, 2013). In this study, doses on contralateral and ipsilateral parotid were reduced on all 22 patients. However CT-on-rails are not available in most clinics. In contrast, CBCT used for patient alignment are common in clinics, and CBCT images are frequently taken during the treatment course. Thus, using CBCT in ART is more practical and efficient than CT-on-rails. Unlike CT, CBCT images contain many artefacts and inaccurate Hounsfield Unit (HU) values, affecting segmentation and dose calculation accuracy. Therefore, CBCT correction needs to be made before use in ART.

Popular methods on CBCT scatter correction include the analytical modeling methods, Monte Carlo (MC) simulation, CT-prior-based methods, histogram matching, and learning-based methods. Analytical modeling methods try to approximate scatter distribution in projection data by assuming that the scatter signal is a convolution function of primary signal and scatter kernel (Naimuddin *et al.*, 1987). The MC simulation-based methods (Xu *et al.*, 2015) are more robust than the analytical method because it simulates photon transport in a flexible yet rigorous way. The prior-CT-based methods exploit prior information obtained from DIR of CT to CBCT (Zöllner *et al.*, 2017). CBCT HU values can also be converted to CT HU values by matching the histograms of CBCT with histograms of CT for each slice via linear scaling (Abe *et al.*, 2017). A regression forest model was trained to correct CBCT images through patient-specific anatomical features extracted from aligned CT and CBCT images (Lei *et al.*, 2018). These methods have achieved limited success and there is still room for improvement. The last three methods all need paired datasets, meaning that CBCT as input and CT as label must have the same exact anatomy. However, even CBCT and CT images acquired on the same day from the same patient would still have slightly anatomical differences, affecting accuracy of simple mapping from CBCT to CT. The current most common practice is to deform planning CT (pCT) though DIR to CBCT anatomy and then to use the deformed planning CT (dpCT) for dose calculation. Thus dpCT keeps HU accuracy as CT. However, the anatomical accuracy of dpCT could be an issue and DIR could yield incorrect contours due to more pronounced anatomical changes and reduced soft-tissue contrast (Kurz *et al.*, 2016). To overcome problems demonstrated above, we would like to generate synthesized CT (sCT) images that have CT's HU values and CBCT's anatomies directly from CBCT images without using paired CT and CBCT images for training.

Deep learning has become a general-purpose solution for image-to-image translation. Hand-engineered mapping functions that would traditionally require complicated formulations are no longer needed (Isola *et al.*, 2017). Therefore, deep learning is a new promising method for translating CBCT to CT images without knowing their mapping functions. One way of using deep learning methods for image-to-image translation is supervised training with paired images. Another way is unsupervised training with unpaired images, like generative adversarial networks (GANs) (Goodfellow *et al.*, 2014), CycleGAN (Zhu *et al.*, 2017), and Triangle-GAN (Gan *et al.*, 2017).

In this paper, our goal is to synthesize CT images from CBCT images that have CT image quality while keeping CBCT anatomy. For supervised learning, it is very difficult to acquire paired CT and CBCT images with exactly matching anatomy. Because of the limitation described above, we applied

CycleGAN, an unsupervised training method to directly convert CBCT to CT-like images. CycleGAN model was used to learn translation functions from a source domain CBCT to a target domain CT. However, in the absence of paired images, training with an inverse mapping from CT to CBCT was coupled at the same time and a cycle consistency loss was introduced to constrain the mapping. This is the first time using unpaired CBCT and CT dataset to train a model which overcomes the limitation of lacking paired CBCT and CT dataset in reality. Unlike DIR, the generated sCT images should keep the exactly same anatomy as CBCT, while having pCT's HU accuracy. We also conducted a deformed phantom study to compare the performance of CycleGAN model against the DIR method in term of anatomical accuracy. We also compared the performance of CycleGAN with two other deep learning models, i.e., deep convolutional generative adversarial network (DCGAN) and progressive growing of GANs, for this particular problem.

## 2. Materials and Methods

### 2.1. Patient data

Data from 13 H&N patients were used for training and validation, while data from 4 patients were used for testing. For each patient, CBCT and CT images were not always acquired on the same day. CBCT images have a resolution of $0.51 \times 0.51 \times 1.99$ mm$^3$ and dimensions of $512 \times 512$ on axial slices. Original CT images have a resolution of $1.17 \times 1.17 \times 3.00$ mm$^3$ and dimensions of $512 \times 512$ on axial slices. CT images in the training dataset are preprocessed by resampling to the same resolution as CBCT. Then, they are aligned with CBCT images and finally cropped to images with dimensions of $512 \times 512$. During training, the CBCT and CT datasets were shuffled at each epoch so that the correspondence to patients is removed. For testing, the ground truth CT images used to benchmark sCT images are generated by deforming pCT to CBCT for each patient. This probably the best one can do since it is very difficult, if not impossible, to acquire CT images with exact anatomy as CBCT images for a patient. By benchmarking sCT against dpCT, we examine the HU value accuracy, not the anatomical accuracy, which will be examined by the phantom study. Eighty axial slices were selected from CBCT and CT images for each patient. The training dataset consisted of 960 CT and CBCT slices, and the validation dataset consisted of 80 CT and CBCT slices. The testing dataset consisted of 320 dpCT and CBCT slices. Both CT and CBCT datasets were normalized to (-1, 1) range for training, validation, and testing.

### 2.2. Overview of CycleGAN

The CycleGAN model is shown in figure 1. Like the other GAN models, CycleGAN includes generators and discriminators that compete against one another until they reach an optimum. Discriminators are used to distinguish real from fake images, while generators are used to generate images that can trick discriminators to believe that the generated fake images are real. Two generators and two discriminators are included in CycleGAN. GeneratorA ($G_A$) is used to generate CT from CBCT and GeneratorB ($G_B$) is used to generate CBCT from CT. DiscriminatorA ($D_A$) is used to distinguish pCT from sCT and DiscriminatorB ($D_B$) is used to distinguish real CBCT from synthesized CBCT (sCBCT). Two cycles are included in CycleGAN. In the first cycle, CBCT is used as input to $G_A$, which generates sCT. Then $G_B$ takes sCT as input and generates cycle CBCT, which is supposed to be equal to CBCT. Meanwhile, $D_A$ identifies real and sCT images. The pCT label is 1 and the sCT label is 0. The second cycle starts with CT as input to $G_B$ which generates sCBCT. Then, $G_A$ takes sCBCT as input and generates cycle CT, which is supposed to be equal to CT. $D_B$ identifies real and sCBCT images simultaneously. The real CBCT label is 1 and the sCBCT label is 0.

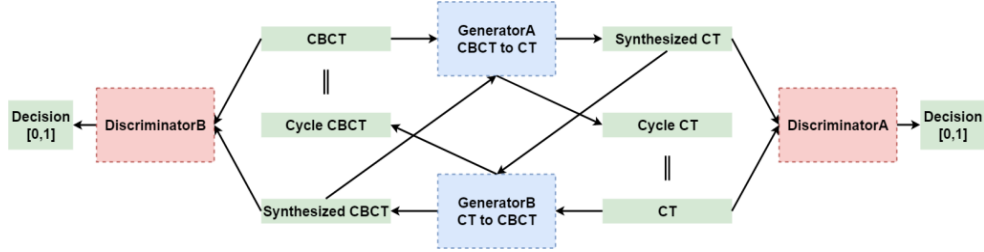

**Figure 1.** The CycleGAN architecture is used to convert CBCT images to sCT images.

### 2.3. Generators and Discriminators of CycleGAN

All models are trained with Adam optimization with a mini-batch size of 2. All weights are initialized from random normal initializer with a mean of 0 and a standard deviation of 0.02. Instance normalization is used after convolution, and LeakyReLU with slope 0.2 is used after instance normalization in all models for fast convergence (Wang *et al.*, 2018). The learning rate is set to 0.0002. We set the momentum term $\beta_1$ to 0.5 to stabilize training (Radford *et al.*, 2015). We use U-Net architecture for generators (figure 2) and 142 × 142 patchGAN for discriminators (figure 3). A 70 × 70 patchGAN was used in image-to-image translation for 256 × 256 image resolution (Isola *et al.*, 2017) because the full ImageGAN not only does not improve the visual quality of the results, but also may even generate lower quality. We choose a 142 × 142 patchGAN as our discriminators for our 512 × 512 image resolution. 142 × 142 is receptive field which is the same size as the input region to the network that contributes to layers activation. The last layer activation function in generators is the Tanh activation function, and the last layer activation function in discriminators is the linear activation function.

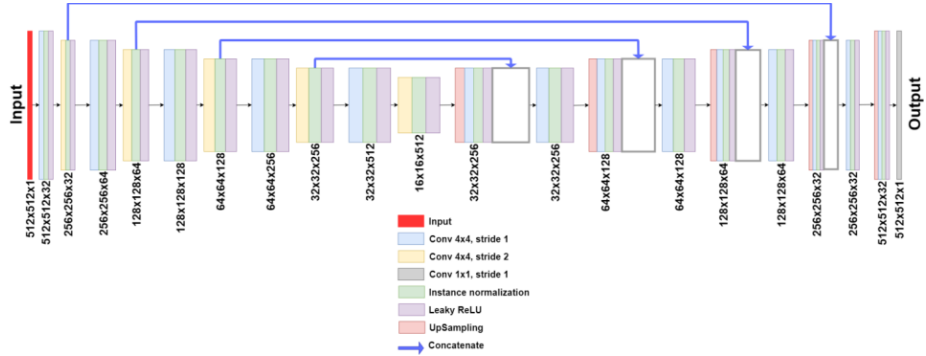

**Figure 2.** U-Net architecture is used for Generators in CycleGAN. The input data size is 512 × 512 × 1 and the output data size is 512 × 512 × 1; the first two numbers represent resolutions and the third number represents channels.

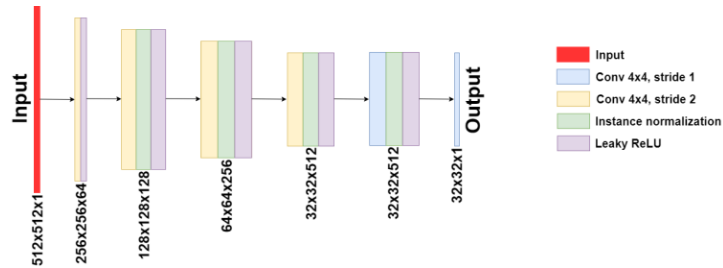

**Figure 3.** The 142 × 142 patchGAN is used for discriminators in CycleGAN. The input data size is 512 × 512 × 1 and the output data size is 32 × 32 × 1; the first two numbers represent resolutions and the third number represents channels.

### 2.4. Loss Function of CycleGAN

As shown in figure 1, CycleGAN includes two mapping functions: mapping from CBCT to CT and mapping from CT to CBCT. Additionally, two discriminators, $D_A$ and $D_B$, are also included in CycleGAN. $D_A$ aims to distinguish pCT from fake CT and $D_B$ aims to distinguish real CBCT from fake CBCT. Our objective includes three types of terms: adversarial loss for mapping the distribution of the generated images to the distribution of the target images; cycle consistency loss to prevent two mappings from contradicting each other; and identity mapping loss to help preserve the gray level of the target images on generated images (Zhu *et al.*, 2017).

Adversarial loss is reflected in both generators and discriminators. $D_A$ aims to classify sCT with label 0 from pCT with label 1, while $G_A$ aims to generate sCT with a label approaching the pCT label. Therefore $D_A$ tries to minimize $\mathcal{L}_{GAN-DA}$,

$$\mathcal{L}_{GAN-DA} = \frac{1}{m}\sum_{i=1}^{m}\frac{(1-D_A(CT_i))^2 + D_A(G_A(CBCT_i))^2}{2}, \quad (1)$$

while $G_A$ tries to minimize $\mathcal{L}_{GAN-GA}$,

$$\mathcal{L}_{GAN-GA} = \frac{1}{m}\sum_{i=1}^{m}(1 - D_A(G_A(CBCT_i)))^2. \quad (2)$$

Similarly, $D_B$ classifies sCBCT with label 0 from real CBCT with label 1. $G_B$ generates sCBCT with a label approaching the real CBCT label. Therefore $D_B$ tries to minimize $\mathcal{L}_{GAN-DB}$,

$$\mathcal{L}_{GAN-DB} = \frac{1}{m}\sum_{i=1}^{m}\frac{(1-D_B(CBCT_i))^2 + D_B(G_B(CT_i))^2}{2}, \quad (3)$$

while $G_B$ tries to minimize $\mathcal{L}_{GAN-GB}$,

$$\mathcal{L}_{GAN-GB} = \frac{1}{m}\sum_{i=1}^{m}(1 - D_B(G_B(CT_i)))^2. \quad (4)$$

A network can still map images from one to multiple domains that share the same distribution features with adversarial loss only. Thus, to further reduce the mapping functions, cycle consistency loss is needed in this task because it minimizes the difference between CT and cycle CT, CBCT and cycle CBCT. The cycle consistency losses for two cycles are

$$\mathcal{L}_{Cycle-CT} = \frac{1}{m}\sum_{i=1}^{m}|G_A(G_B(CT_i)) - CT_i|, \quad (5)$$

and

$$\mathcal{L}_{Cycle-CBCT} = \frac{1}{m}\sum_{i=1}^{m}|G_B(G_A(CBCT_i)) - CBCT_i|, \quad (6)$$

To further preserve HU values between pCT and sCT, real CBCT, and sCBCT, we add identity mapping loss in the loss function. $G_A$ supposedly generates sCT from CBCT. However, if pCT images are input to $G_A$, the output is supposed to be pCT too and vice versa. Therefore, the identity mapping loss for CT and CBCT are expressed as follows:

$$\mathcal{L}_{Identity-CT} = \frac{1}{m}\sum_{i=1}^{m}|G_A(CT_i) - CT_i|, \tag{7}$$

$$\mathcal{L}_{Identity-CBCT} = \frac{1}{m}\sum_{i=1}^{m}|G_B(CBCT_i) - CBCT_i|. \tag{8}$$

Thus, combining all the losses above, the full objective for two generators is

$$\mathcal{L}_G = \mathcal{L}_{GAN-GA} + \mathcal{L}_{GAN-GB} + \alpha \times (\mathcal{L}_{Cycle-CT} + \mathcal{L}_{Cycle-CBCT}) + \beta \times (\mathcal{L}_{Identity-CT} + \mathcal{L}_{Identity-CBCT}), \tag{9}$$

where $\alpha$ is 10 and $\beta$ is 5.

The full objective for two discriminators is

$$\mathcal{L}_D = \mathcal{L}_{GAN-DA} + \mathcal{L}_{GAN-DB}. \tag{10}$$

### 2.5 Evaluation

We have two hypotheses to test. Hypothesis 1: sCT is more accurate than CBCT in terms of HU value accuracy; Hypothesis 2: sCT is more accurate than dpCT in terms of anatomical accuracy.

To test Hypothesis 1, we use dpCT as reference, which is the best we can do for real patient data. 320 axial slices from four patients were used for testing, with the dimension of $512 \times 512$ for each CT and CBCT images. We deformed pCT images to their corresponding CBCT images anatomy to obtain the ground truth data. The dpCT images, generated using DIR in Velocity Oncology Imaging Informatics system (Varian Medical Systems, Inc.), have close anatomy as CBCT images while their HU values are kept the same as pCT. We examine image quality visually, perform the linear regression of the HU value scatter plots, compare the HU value line profiles, and compare the Q-Q (quantile-quantile) plots. We also conduct a dosimetric study by casting patients' treatment plans on CBCT, sCT, and dpCT.

To test Hypothesis 2, we perform a phantom study. A deformable H&N phantom (Graves *et al.*, 2015) was used to compare sCT images from the CycleGAN model and dpCT images from the DIR method, in terms of anatomical accuracy. The phantom construction is shown in figure 4. In the phantom, tumor (area1) and parotid gland (area2) are deformable by injecting saline water and moving diode holders on rail. We simulated tumor shrink and anatomy change by injecting 60 ml saline water to the phantom, then drawing 30 ml saline water out from the phantom and adjusting the diode holder positions. CT1 is acquired under 60 ml saline water, simulating pCT. CT2 and CBCT2 are acquired under 30 ml saline water, simulating the re-scanned CT and CBCT on a particular day during the treatment course. sCT is generated by CycleGAN model from CBCT2 and dpCT is generated by deforming CT1 to CBCT2 anatomy using the DIR method in Velocity Oncology Imaging Informatics System (Varian Medical Systems, Inc.). CT2 serves as the ground truth for quantitative evaluation.

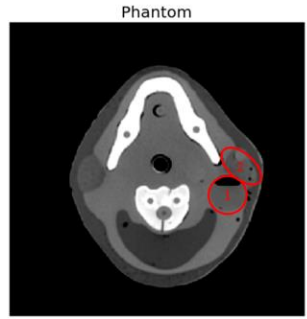

**Figure 4.** Illustration of a deformable H&N phantom. Area 1 represents tumor and area 2 represents parotid gland.

## 3. Results

### *3.1. Scatter removal and HU improvement*

Figure 5 shows the testing CBCT images, and the corresponding sCT and dpCT images. One can see that the sCT images generated from CBCT through the CycleGAN method have artifacts greatly reduced while keeping the same anatomy as CBCT. In contrast, the dpCT images have slightly different anatomy distribution.

The linear regression of the scatter plots of all testing data are shown in figure 6. If two datasets come from the same distribution, their linear regression plot should fall on the 45-degree reference line. The sCT and dpCT regression line is close to the 45-degree reference line indicating that those two datasets are very similar and there is a great improvement of sCT over CBCT in terms of HU value accuracy.

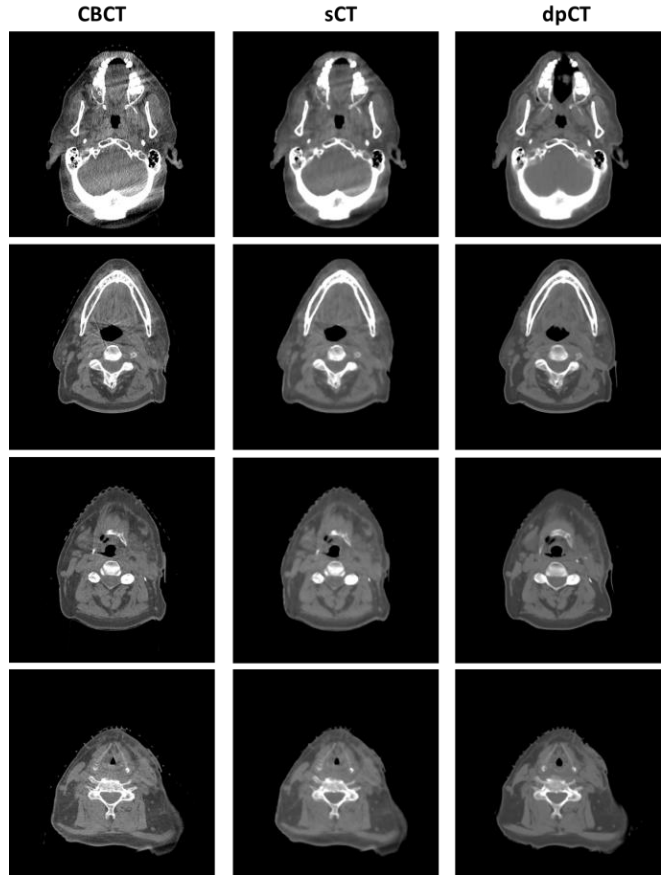

**Figure 5.** Images for four testing patients, one for each row. First column: CBCT images; Second column: sCT images; Third column: dpCT images. The display window is [-300,500] HU.

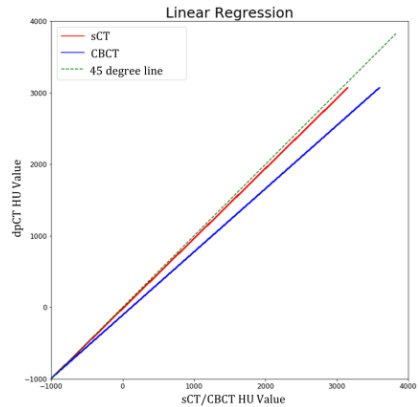

**Figure 6.** The linear regression of the scatter plot of sCT versus dpCT (red line) falls closely onto the 45-degree reference line, while the linear regression of the scatter plot of CBCT versus dpCT (blue line) falls away from the 45-degree reference line. The green dash line is the 45-degree reference line. The slope of red line is 0.98 and intercept is -19.2, while for blue line the slope is 0.88 and intercept is -110.6.

We use four similarity measures: mean absolute error (MAE), root mean squared error (RMSE), structural similarity index (SSIM) and peak signal-to-noise ratio (PSNR) to evaluate HU accuracy of all testing sCT images and CBCT images compared with dpCT images quantitatively. The results are shown in Table 1. MAE and RMSE of sCT using CycleGAN are decreased to 27.98 HU and 86.71 HU from 71.78 HU and 167.22 HU of CBCT. SSIM and PSNR of sCT are increased to 0.85 and 30.67 from 0.77 and 25.22 of CBCT against CT.

**Table 1.** Similarity measures between CBCT and dpCT images, sCT and dpCT images of CycleGAN method with all the testing patient dataset.

|  | MAE (HU) | RMSE (HU) | SSIM | PSNR |
|---|---|---|---|---|
| **CBCT Against dpCT** | 71.78 | 167.22 | 0.77 | 25.22 |
| **sCT Against dpCT** | 27.98 | 86.71 | 0.85 | 30.67 |

We also plot typical line profiles on one sample of the testing data (figure 7). In the upper line profile, which passes through soft tissue and bone areas, sCT HU values are corrected to CT HU values. In the lower line profile, which passes only through the soft tissue area, CBCT HU values are noisy while sCT HU values are smoothed and corrected to CT HU values. CBCT HU values have been corrected to the CT HU values on sCT regardless of bone or soft tissue areas.

We also applied the Q-Q plot to evaluate an example of testing dataset: dpCT, CBCT, and sCT using the cycle-GAN method (figure 8). The points of dpCT-sCT Q-Q plots falling along the 45-degree reference line regardless of soft tissue or bone areas indicate that sCT has the same distribution as dpCT. Thus CycleGAN can effectively generate CT-like images from CBCT based on both visual and quantitative evaluation results.

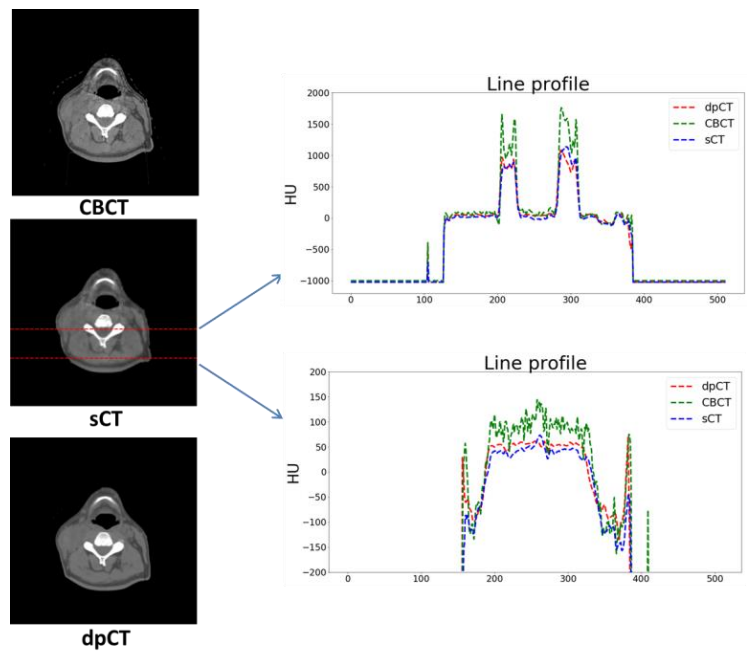

**Figure 7.** The left row illustrates an example of CBCT image, sCT image synthesized from CBCT by CycleGAN, and dpCT image deformed from planning CT, all from the testing data. Their display window is [-300,500] HU. The right plots show HU profiles of the two red dashed lines in the left image. The upper line profile passes through bone and soft tissue. The lower line only passes through soft tissue only.

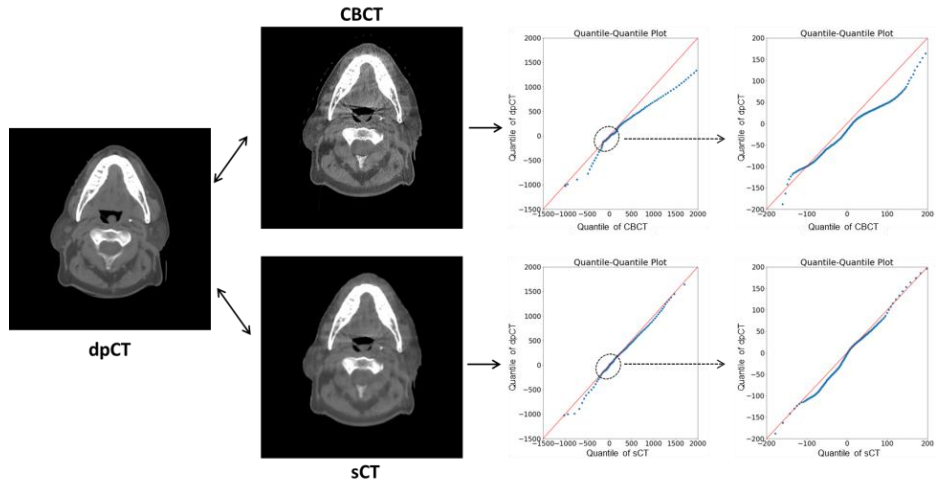

**Figure 8.** The first row shows dpCT-CBCT $Q - Q$ plots, ranging from -1500 HU to 2000 HU (including bones) and -200 HU to 200 HU (soft tissue). The second row shows dpCT-sCT $Q - Q$ plots, ranging from -1500 HU to 2000 HU and -200 HU to 200 HU. The points on dpCT-sCT $Q - Q$ plots fall on the 45-degree reference red line, indicating that sCT and dpCT data sets come from a common distribution. Display window is [-300,500] HU.

### *3.2 Dose distribution comparison*

Figure 9 shows dose calculation results of photon beam treatment plans using CBCT, sCT and dpCT images of four testing patients. Visually, the dose distributions of the sCT are found to be closer to those of the dpCT, while dose distributions of the CBCT show obvious discrepancies from those of the dpCT. Quantitatively, the results of 3D gamma analysis are shown in table 2. Dose distributions on sCT have higher gamma pass rates with mean value of 97.40% compared to dose distributions on CBCT with mean value of 95.42% under 1 mm/1% gamma criteria.

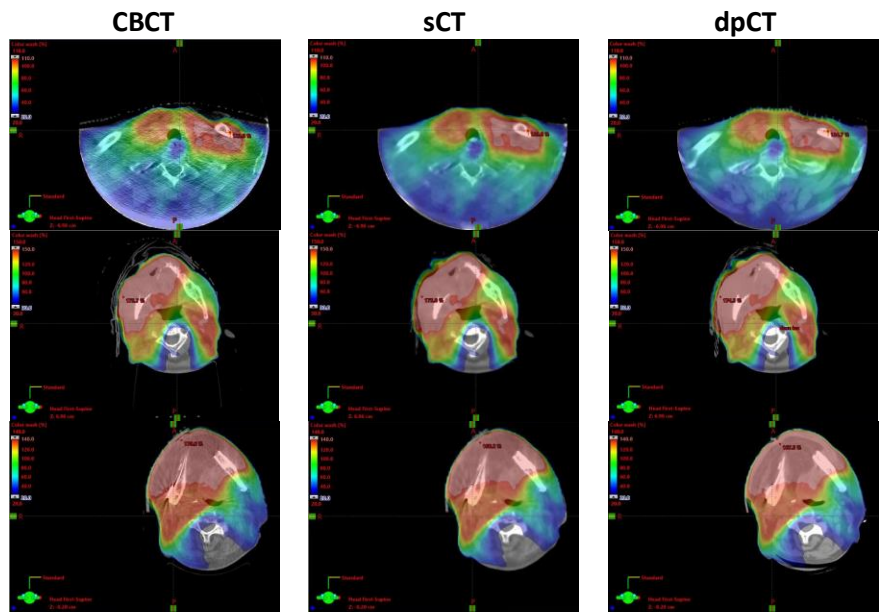

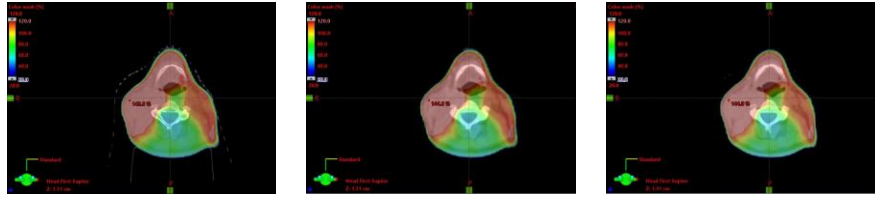

**Figure 9.** Dose distributions of photon beam treatment plans using CBCT (left column), sCT (middle column) and dpCT (right column). Each row represents a different testing patient.

**Table 2.** 3D gamma analysis of dose distributions on CBCT and sCT compared with those on dpCT for photon beam treatment plans on four testing patients. The numbers are passing rate of gamma index.

|  | Gamma criteria: 2 mm/2% | | Gamma criteria: 1 mm/1% | |
| --- | --- | --- | --- | --- |
|  | **CBCT** | **sCT** | **CBCT** | **sCT** |
| Patient1 | 99.58% | 100% | 97.30% | 99.92% |
| Patient2 | 93.36% | 95.42% | 88.78% | 90.18% |
| Patient3 | 99.03% | 99.97% | 97.58% | 99.57% |
| Patient4 | 99.62% | 100% | 98.02% | 99.94% |
| Average | 97.90% | 98.85% | 95.42% | 97.40% |

### 3.3 Phantom study

The images of CT1, CBCT2, sCT, dpCT and CT2 are shown in figure 10. sCT generated from CBCT2 keeps the same anatomy as CT2, while at the same time removes artifacts from CBCT2. By comparing dpCT against CT2, we can tell that the DIR method doesn't perform very well in terms of capturing the anatomy change in this phantom study. Table 3 shows results of similarity measures of CT1, CBCT2, sCT and dpCT with ground truth CT2. For sCT, MAE of 4.66 HU and RMSE of 16.28 HU are lower than those for dpCT (MAE of 6.98 HU and RMSE of 22.97 HU), while SSIM of 0.95 and PSNR of 32.70 are higher than those for dpCT (SSIM of 0.91 and PSNR of 29.75), showing better performance of the CycleGAN model than the DIR method in this case.

**Table 3.** Similarity measures for CT1, CBCT2, sCT, and dpCT against CT2 in the phantom study.

|  | **CT1** | **CBCT2** | **sCT** | **dpCT** |
| --- | --- | --- | --- | --- |
| **MAE (HU)** | 8.43 | 28.54 | 4.66 | 6.98 |
| **RMSE (HU)** | 28.40 | 59.46 | 16.28 | 22.97 |
| **SSIM** | 0.89 | 0.73 | 0.95 | 0.91 |
| **PSNR** | 27.22 | 21.06 | 32.70 | 29.75 |

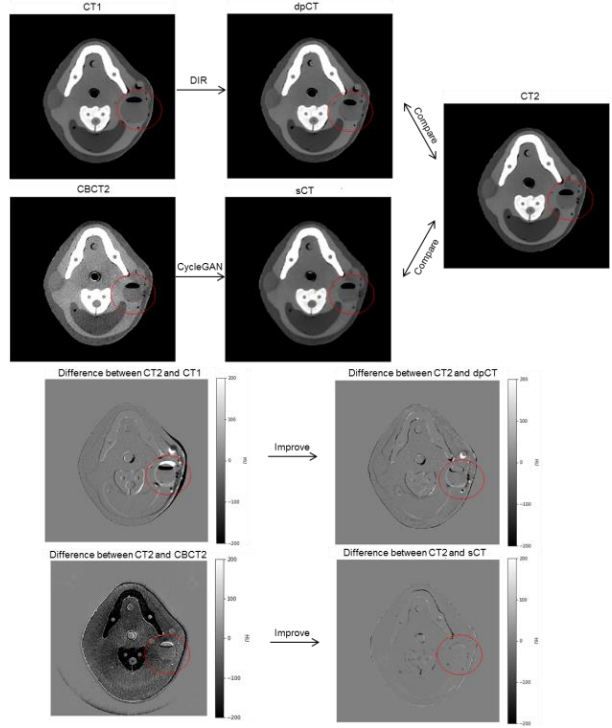

**Figure 10.** CT1 was scanned as the planning CT (pCT) before tumor shrinkage. CT2 and CBCT2 were scanned after tumor shrinkage, as if some treatment fractions have been delivered. sCT was generated using the CycleGAN method from CBCT2, and dpCT was generated using the DIR method from CT1. Their display window is [-150,500] HU. Display window of difference maps is [-200,200] HU.

### 4. Discussion and Conclusions

A CycleGAN-based model has been developed to convert CBCT to CT-like images. In the model, unpaired CBCT and CT images can be used for training to avoid the requirement of paired CBCT and CT datasets that share exactly the same anatomy, which are difficult to acquire in reality. Once training is completed, generating sCT image from CBCT by this model can be accomplished in seconds. Furthermore, through our model most artifacts in CBCT are removed and the HU values are corrected, using dpCT images as reference.

In the real patient study, dpCT images are used as ground truth to evaluate HU accuracy of sCT images since it is difficult to acquire CT images that have exactly the same anatomy as CBCT images. Even though dpCT images have minor anatomical accuracy issues, dpCT images keep the same HU accuracy as real CT images. Therefore it is reasonable to use dpCT to evaluate sCT's HU accuracy. On the other hand, sCT's anatomical accuracy was evaluated with CT images in the phantom study when acquiring CBCT and CT images with exactly same anatomy is possible. It is also possible to compare CycleGAN model and DIR method in the phantom study. The deformable phantom study shows better performance of CycleGAN model over DIR with MAE decrease from 6.98 HU to 4.66 HU and SSIM increase from 0.91 to 0.95.

In this experiment, we also tried another two unsupervised deep learning methods which are deep convolutional generative adversarial network (DCGAN) (Radford *et al.*, 2015) and progressive growing of GANs (Karras *et al.*, 2017) to train the models with the same training and testing patient dataset. Similarity measures show that MAE, RMSE, SSIM and PSNR of sCT against dpCT images by DCGAN are 40.66 HU, 120.56 HU, 0.81 and 28.67, while MAE, RMSE, SSIM and PSNR of sCT against dpCT images by progressive growing of GANs are 36.76 HU, 113.80 HU,0.82 and 28.00. DCGAN removes less artefacts because of its less restrained loss function and progressive growing of GANs tends to generate different anatomy from original CBCT images. In our training and testing dataset, some CBCT images have truncation problems, progressive growing of GANs tries to add new anatomy on the truncated part. However it generates false anatomy which makes it to the worst method in our experiment. It is also very time costly during training since it has different resolutions to train. Compared with these two models, CycleGAN obtained the best image quality visually and quantitatively.

By our CycleGAN model, HU accuracy of sCT images is improved compared with CBCT images, while anatomy of sCT is kept the same with CBCT images. sCT images generated by CycleGAN have similar HU values with dpCT images, while anatomical accuracy by CycleGAN outperforms the DIR method. The sCT images generated by CycleGAN with less artifact and corrected HU values can be used for more accurate segmentation and dose calculation in ART. For future work, we need to improve the performance on metal artifacts and exponential edge gradient effect. We also need to solve the image truncation problem caused by the small field of view, in order to use sCT for ART planning. The training and testing of the model for other anatomical sites are our natural next step as well.

**Acknowledgement**

We would like to thank the financial support of Cancer Prevention and Research Institute of Texas (CPRIT) through the grant IIRA RP150485.